\pdfoutput=1
\documentclass[conference,10pt]{IEEEtran}
\usepackage{amsmath}
\usepackage{amssymb}
\usepackage{bm}
\usepackage{PaperStyle}

\begin{document}
\title{Inferring Sparsity: Compressed Sensing using Generalized Restricted Boltzmann Machines}

\author{\IEEEauthorblockN{Eric W. Tramel\IEEEauthorrefmark{2},
                          Andre Manoel\IEEEauthorrefmark{2},
                          Francesco Caltagirone\IEEEauthorrefmark{3}, 
                          Marylou Gabri\'e\IEEEauthorrefmark{2} and
                          Florent Krzakala\IEEEauthorrefmark{2}\IEEEauthorrefmark{4}}
\IEEEauthorblockA{\IEEEauthorrefmark{2}\small
                  Laboratoire de Physique Statistique (CNRS UMR-8550),\\
                  \'Ecole Normale Sup\'erieure, PSL Research University, 
                  24 rue Lhomond, 75005 Paris, France}
\IEEEauthorblockA{\IEEEauthorrefmark{4}\small
                  Universit\'e Pierre et Marie Curie, Sorbonne Universit\'es, 75005 Paris, France}
\IEEEauthorblockA{\IEEEauthorrefmark{3}\small
                  INRIA Paris, 2 rue Simone Iff, 75012 Paris, France}
\vspace{-0.5cm}}

\maketitle
\begin{abstract}
In this work, we consider compressed sensing reconstruction from 
$M$ measurements of $K$-sparse structured 
signals which do not possess a writable correlation model. 
Assuming that a generative statistical model, such as a Boltzmann machine, can
be trained in an unsupervised manner on example signals, we demonstrate how this
signal model can be used within a Bayesian framework of signal reconstruction. 
By deriving a message-passing inference for general distribution restricted 
Boltzmann machines, we are able to integrate these inferred signal models into
approximate message passing for compressed sensing reconstruction. Finally, we 
show for the MNIST dataset that this approach can be very effective, even for
$M < K$.
\end{abstract}

\IEEEpeerreviewmaketitle

\section{Introduction}
Over the past decade, the study of compressed sensing (CS) 
\cite{CW2008,CR2005a,Don2006} has lead to many significant developments in the
field of signal processing including novel sub-Nyquist sampling strategies
\cite{TLW2006,MED2011} and a veritable explosion of work in sparse approximation 
and representation \cite{ZXY2015}. 
The core problem in CS is the reconstruction of a sparse signal of 
dimensionality $N$ from a set of $M$ noisy observations for $M\ll N$.
Here, a \emph{sparse} signal is defined as one which possesses many fewer 
non-zero coefficients, $K$, than its ambient dimensionality, $K\ll N$. 
The theoretical foundations of CS recovery conditions are built upon the concept
of \emph{support identification}, 
finding the locations of these non-zero coefficients. 
Tackling the reconstruction directly is combinatorially hard, requiring a search over 
all ${N\choose K}$ possible support patterns for the one which best matches the
given observations. In the noiseless setting, if such a search were possible,
we require at least $M=K$ for the on-support values to be perfectly estimated.
As we do not assume any particular distribution on the $K$ non-zero values, 
this requirement follows directly from linear algebra.  
It was shown in \cite{CR2005a} that a convex relaxation from
a strict requirement on $K$-sparsity to an $\ell_1$ regularization allows for 
efficient reconstruction, but requires $M \gtrsim K\log N$ for exact reconstruction.
Greedy approaches \cite{DTD2006,DGN2008} retain $K$-sparsity, solving the
reconstruction problem by iterating between support identification and 
on-support signal estimation. However, these robust and conceptually simple 
techniques come at the cost of increased $M$ for exact reconstruction.

Since the identification of sharp thresholds between successful and 
unsuccessful signal recovery as a function of 
$\alpha=M/N$ and $\rho=K/N$ \cite{DT2006}, these 
\emph{phase transitions} in the space of possible CS reconstruction problems 
have been used as a tool for comparing different CS recovery strategies. In 
\cite{BSB2010,Ran2010}, it was shown that using sum-product belief propagation (BP) 
in conjunction with a two-mode Gauss-Bernoulli sparse signal prior
provided a much more favorable phase transition, as compared to $\ell_1$ 
minimization, for $K$-sparse signal reconstruction.
In \cite{DMM2009}, approximate message passing (AMP) was proposed as an 
efficient alternative to BP
and was further refined in a series 
of papers \cite{DMM2010b,Ran2011,KMS2012}.

While these Bayesian techniques have had a significant impact in improving the 
lower bound on $M$ for exact reconstruction, 
the $M=K$ lower bound on general recoverability can only be approached
in the case that
directly designing the sampling strategy \cite{KrzakalaPRX2012} is possible.
As such designs are often not
practically achievable, decreasing the requirement on $M$ necessitates the use 
of more informative signal priors, e.g. a prior which leverages
\emph{a priori} known correlations for the signal class of interest.
The works \cite{dremeau2012boltzmann,tdk2016} 
sought to model support correlation directly by leveraging the abstraction power
of latent variable models via Boltzmann machines trained on examples of signal
support. While these techniques demonstrated significant improvements in 
recovery performance for sparse signals, they are still fundamentally bound by
the $M=K$ transition. The unification of machine learning approaches with
CS was also addressed in the recent works \cite{KLT2016,MPB2015,KM2015}, 
in which feed-forward deep neural network structures are used to aid CS 
signal reconstruction. 

In this work, we investigate the possibility of modeling both signal and
support, as in \cite{schniter2010turbo,rangan2011hybrid}, using however a
trained latent variable model as prior for the AMP reconstruction. For this, 
we turn to real-valued restricted Boltzmann machines (RBMs). 
In order to utilize real-valued RBMs within the AMP framework, we propose an
extended mean-field approximation similar in nature to \cite{tdk2016,GTZ2015}.
However, we extend this approximation to the case of \emph{general distributions}
on both hidden and visible units of the RBM, allowing us to model sparse
signals directly. Given this trained RBM, we propose a 
CS reconstruction algorithm which amounts to two nested inference problems, one on 
the CS observation-matching problem, and the other on the RBM
model. In our results, we show that this technique can provide good
reconstructions even for $M<K$, 
as the RBM signal prior not only models the support correlation structure, but
the joint distribution of the on-support values, as well.

\section{Background}


In the CS problem,
we wish
to recover some unknown $K$-sparse signal $\x\in\real^{N}$ given a set of observations
$\y\in\real^{M}$, $M\ll N$, generated by
$\y = \F\x + \w$ where   
$\w \sim \normal{\mathbf{0}}{\Delta \mathbf{I}}$
and the matrix $\F\in\real^{M\times N}$ is a random projection operator.
While a number of different output channels of the form
$\y = g(\F\x)$ could be conceived \cite{Ran2011}, 
for clarity we focus on the case of an additive white Gaussian noise 
(AWGN) channel.

Following the Bayesian approach to signal reconstruction, 
we will focus on estimation techniques involving the posterior distribution
\begin{equation}
    P(\x | \F, \y) =\frac{ e^{-\frac{1}{2\Delta}\| \y - \F\x \|_2^2} \, P_0 (\x)}
    {\integ{\x} e^{-\frac{1}{2\Delta}\| \y - \F\x \|_2^2} \, P_0 (\x)}.
    \label{eq:true_posterior}
\end{equation}
Even if computing the moments of \eqref{eq:true_posterior} is intractable 
for some $P_0(\x)$,
\cite{DMM2009,KMS2012} show that
the minimum mean-square-error (MMSE) estimator, $\hat{\x}_\text{MMSE} (\F,
\y) = \integ{\x} \x \, P(\x|\F, \y)$, can be computed extremely efficiently
using loopy BP or AMP whenever $P_0(\x)$ is fully factorized.

The AMP algorithm \cite{DMM2009,KMS2012,Ran2011},
under stricter  requirements of \iid random $\F$, 
provides at each step of its iteration, 
an approximation to the posterior of the form
\begin{equation}
    Q (\x | \bm{A}, \bm{B}) = \frac{1}{\Z (\bm{A}, \bm{B})} \,
    P_0 (\x) \, e^{-\frac{1}{2} \sum_i A_i x_i^2 + \sum_i B_i x_i},
    \label{eq:amp-approx}
\end{equation}
where $\mathcal{Z} (\bm{A}, \bm{B})$ is a normalization, and $\bm{A}$
and $\bm{B}$ are quantities obtained by iterating the following AMP equations,
\begin{align}    
    V_{m}^{\rm(t+1)} &= 
            \sum_i F_{mi}^2 c_i^{\rm(t)}, \label{eq:ampV}\\
    \omega_{m}^{\rm(t+1)} &= 
            \sum_i F_{mi} a_i^{\rm(t)} - V_m^{\rm(t+1)}\frac{y_m - \omega_m^{\rm(t)}}{\Delta + V_m^{\rm(t)}}, \label{eq:ampw}\\
    A_i^{\rm(t+1)} &= 
            \sum_m \frac{F_{mi}^2}{\Delta + V_m^{\rm(t+1)} }, \label{eq:ampA}\\
    B_i^{\rm(t+1)} &= 
            A_i^{\rm(t+1)}a_i^{\rm(t)} + \sum_m F_{mi} \, \frac{y_m - \omega_m^{\rm(t+1)}}{\Delta + V_m^{\rm(t+1)}},
            \label{eq:ampB}
\end{align}
where $\veca^{(\rm t)}$ and $\vecc^{(\rm t)}$ are the
mean and variance of $Q (\x | \bm{A}^{(\rm t)}, \bm{B}^{(\rm t)})$, 
which after the convergence of the algorithm provide an approximation 
of the mean and variance of \eqref{eq:true_posterior}.
These moments, given a computable $\Z
(\bm{A}, \bm{B}) = \int {\rm d}\x \, P_0 (\x) \, e^{-\frac{1}{2} \sum_i A_i
x_i^2 + \sum_i B_i x_i}$, are easily obtainable from
\begin{equation} 
    a_i \defas 
    \frac{\partial\ln \Z(\bm{A}, \bm{B})}{\partial B_i},
    ~~
    c_i \defas \frac{\partial^2\ln \Z(\bm{A}, \bm{B})}{\partial B_i^2}.
    \label{eq:fafc}
\end{equation}
In particular, whenever the prior distribution is fully factorized, $P_0
(\x) = \prod_i P_0 (x_i)$, evaluating $\Z(\bm{A}, \bm{B})$ amounts to solving
$N$ independent one-dimensional integrals. 
In contrast, for general $P_0(\x)$, the normalization is intractable.
In this case one must resort to further approximations.

In what follows, we use an RBM \cite{Smo1986} to model the signal's prior 
distribution jointly with a set of latent, or \emph{hidden}, 
variables $\h$,
\begin{equation}
    P_0 (\x, \h | \bm{\theta}) = \frac{1}{Z(\bm{\theta})}~e^{\x^T \W
        \h} \prod_i P_0 (x_i | \bm{\theta}_{\x}) \prod_{\mu} P_0 (h_{\mu} |
    \bm{\theta}_{\h}),
    \label{eq:rbmprior}
\end{equation}
and the parameters $\bm{\theta} = \{\W, \bm{\theta}_{\x},
\bm{\theta}_{\h} \}$ can be obtained by training the RBM over a
set of examples \cite{Hin2002,GTZ2015}. This construction defines
a \emph{generalized} RBM (GRBM), in the sense
that the visible and hidden variables are not strictly binary and may possess
any distribution. Using a GRBM as a signal prior, the normalization of 
\eqref{eq:amp-approx}, 
\begin{align}
    \Z (\bm{A}, \bm{B}) &= \frac{1}{Z (\bm{\theta})}~\int \left[ \prod_i {\rm
        d}x_i \, P_0 (x_i) \, e^{-\frac{A_i}{2} x_i^2 + B_i x_i} \right]
    \, \times \notag \\
    &\kern5em \int \left[ \prod_{\mu} {\rm d}h_{\mu} \, P_0 (h_{\mu}) \right]
        e^{\x^T \W \h},
    \label{eq:partition}
\end{align}
is no longer factorized or tractable, thus
requiring some approximation to calculate the necessary moments of 
$Q$. In the next section we introduce a message-passing algorithm to
evaluate (\ref{eq:partition}) and estimate these moments.

\begin{figure}
    \centering
    \includegraphics[width=0.42\textwidth]{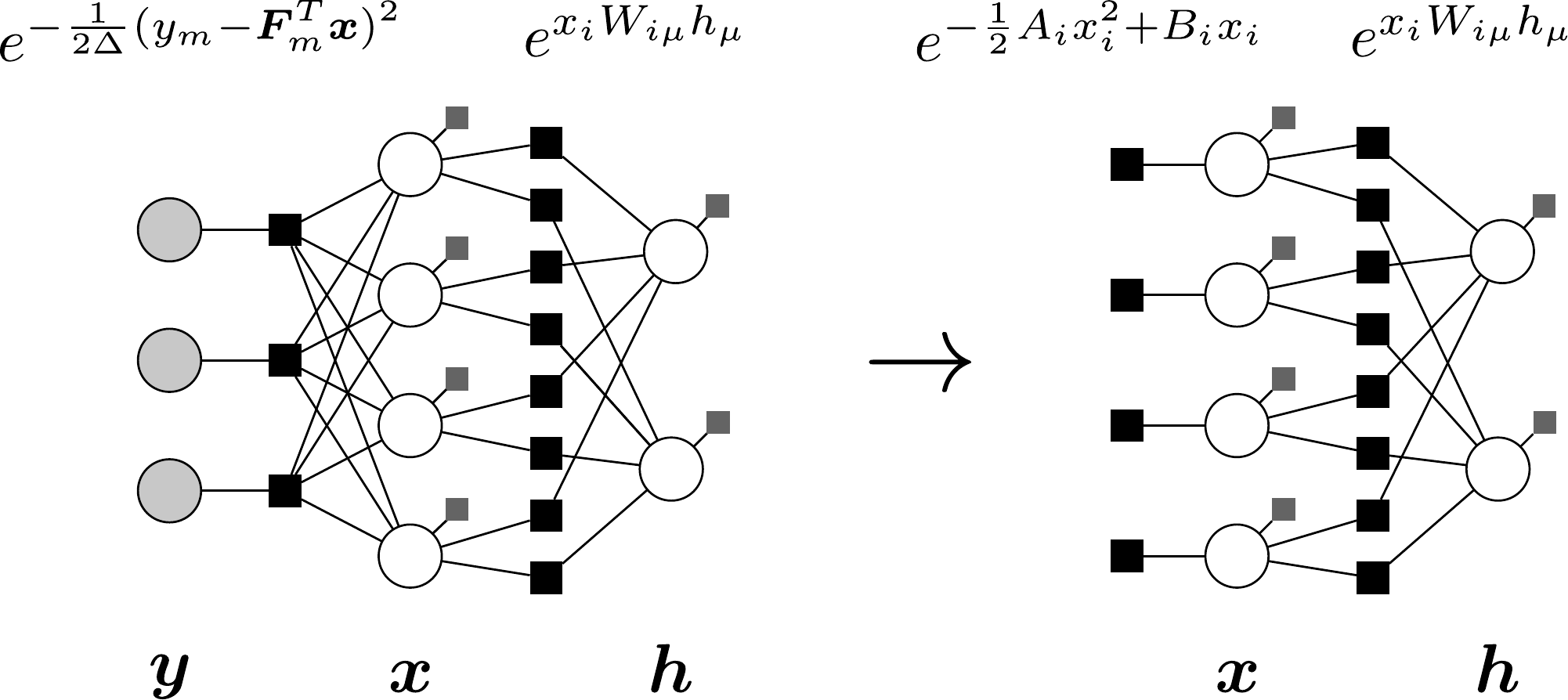}
    \caption{\emph{Left:} Factor graph representation of the posterior 
    \eqref{eq:true_posterior} given the RBM signal prior
    \eqref{eq:rbmprior}. Circles and squares represent variables and factors,
    respectively. Light gray circles are the observed signal measurements.
    Black factors are the factors induced by linear operators $\F$ and $\W$.
    Light gray factors represent prior distributions influencing their adjoining
    variables. \emph{Right:} Factor graph for the approximating posterior
    \eqref{eq:amp-approx}. Factors on the left represent local
    effective potentials provided by AMP at each of its iterations.}
    \label{fig:factor}
\end{figure}

\section{TAP Approximation of a General-case RBM}
In order to use the RBM model (\ref{eq:rbmprior}) as a prior within AMP, we
must perform inference over the RBM
graphical model given on the right of Fig. \ref{fig:factor}.
Specifically, we construct a message-passing scheme between the factors and the
hidden and visible variables of the RBM. Loopy BP has been 
considered in the context of inference on RBMs in \cite{HT2015}, where
the authors assume a Bernoulli distribution on the hidden variables and
rewrite them as factors. 
In contrast, we investigate a more general
setting with arbitrary distributions on both the hidden and visible variables. 
Since all factors have degree 2, using BP 
we can write the messages from variable to variable,
\begin{align}
\psi\itm(x_i) 
  &=
  \frac{1}{\Z\itm} P_0(x_i) e^{-\frac{1}{2}\Aiamp x_i^2 + \Biamp x_i}\notag\\
  &\quad\quad\times\prod_{\nu\in\partial i / \mu}
    \integ{\hv}\psi\nti(\hv) e^{x_i\Wiv\hv},
\end{align}
\begin{align}
\psi{\mti}(\hu) 
  &=
    \frac{1}{\Z\mti} P_0(\hu) \notag\\
  &\quad\quad 
    \times\prod_{j\in\partial\mu / i} \integ{x_j}\psi\jtm(x_j) e^{x_j\Wju\hu},
\end{align}
where we index hidden units by $\mu$, $\nu$ and visible units by $i$, $j$, and
the notation $\partial i / \mu$ represents the set of all edge-connected 
neighbors of variable $i$ \emph{except} $\mu$.
If we assume that the magnitudes of the values of $\W$ are small, then we can
perform an expansion on the messages \cite{KMS2012}.
This is in essence the relaxed BP of
\cite{Ran2010}, transitioning from messages of continuous distributions to messages 
parameterized by their first two central moments, denoted by the letters $a$ and
$c$, respectively. Specifically,
\begin{align}
&\integ{\hv}\psi\nti(\hv) e^{x_i\Wiv\hv} \notag\\
  &~= 
    \bexp{x_i\Wiv a^{\rm h}\nti + \frac{1}{2}x_i^2\Wiv^2 c^{\rm h}\nti} + O(\W^3),\\
&\integ{x_j}\psi\jtm(x_j) e^{x_j\Wju\hu} \notag\\
  &~= 
    \bexp{\hu\Wju a^{\rm v}\jtm + \frac{1}{2}\hu^2\Wju^2 c^{\rm v}\jtm} + O(\W^3),
\end{align}
where $O(\W^3)$ represents a vanishing third-order correction to the two-moment
approximation of the message marginalizations which is dependent on the RBM parameters
$\W$. As we are now able to move the product
of incoming messages as a sum into the exponent, we define the following 
intermediate sum variables as in AMP,
\begin{align}
  A^{\rm v}\itm \defas
      -\sum_{\nu\in H / \mu} \Wiv^2 c^{\rm h}\nti,  
  ~&
  B^{\rm v}\itm \defas
      \sum_{\nu\in H / \mu} \Wiv a^{\rm h}\nti,\\
  A^{\rm h}\mti \defas 
      -\sum_{j\in V / i}\Wju^2 c^{\rm v}\jtm,
  ~&
  B^{\rm h}\mti \defas
      \sum_{j\in V / i}\Wju a^{\rm v}\jtm,
\end{align}
where the sets $H$ and $V$ are the set of all hidden and visible
variables, respectively.
From these definitions, we can define the message moments explicitly and 
close the message passing equations on the edges of the RBM factor graph,
\begin{alignat}{2}
  a^{\rm h}\mti &= \fha(A^{\rm h}\mti, B^{\rm h}\mti),
  &&c^{\rm h}\mti = \fhc(A^{\rm h}\mti, B^{\rm h}\mti),\\
  a^{\rm v}\itm &= \fva(\Aiamp + A^{\rm v}\itm,\,&&\Biamp + B^{\rm v}\itm),\\
  c^{\rm v}\itm &= \fvc(\Aiamp + A^{\rm v}\itm,\,&&\Biamp + B^{\rm v}\itm),
\end{alignat}
where the prior-dependent functions for the 
visible and hidden variables, $(\fva,\fvc)$ and 
$(\fha,\fhc)$ respectively, 
are defined in a fashion similar to \eqref{eq:fafc}, i.e. as the moments
of an approximating distribution similar to \eqref{eq:amp-approx}, 
but using the desired hidden and visible 
distributions in place of $P_0$.
The marginal beliefs at each hidden and
visible variable can be defined by summing over \emph{all} incoming messages. 
One could run this message passing, as stated, until convergence
on the beliefs in order to infer marginal distributions on both the hidden and
visible variables of the RBM. The moments of the marginal distributions on
the visible units, $\avi$ and $\cvi$, would give us exactly the moments
required by the AMP iteration.

However, such a message passing on the edges of the factor graph
can be quite memory and computationally intensive, especially if this inference
occurs nested as an inner loop of AMP.
If we assume that the entries of $\W$ are widely distributed, without any 
particular strong correlations in its structure,
we can construct an algorithm which operates entirely on the beliefs, the
nodes of the factor graph, rather than the messages, the edges. Such an
algorithm is similar in spirit to AMP and also to the
Thouless-Anderson-Palmer (TAP) equations from statistical physics.
We now write these TAP self-consistency equations closed on the
parameters of the marginal beliefs alone,
\begin{alignat}{2}
  \Ahm &= -\sum_{i\in V}\Wiu^2\cvi, \label{eq:tap_AhBh}
  ~
  &&\Bhm = \ahm\Ahm + \sum_{i\in V} \Wiu \avi, \\
  \ahm &= \fha(\Ahm,\Bhm), \label{eq:tap_ahch}
  ~
  &&\chm = \fhc(\Ahm,\Bhm), \\
  \Avi &= -\sum_{\mu\in H} \Wiu^2 \chm, \label{eq:tap_AvBv}
  ~
  &&\Bvi = \avi\Avi + \sum_{\mu\in H} \Wiu \ahm ,\\
  \ai &= \fva(\Aiamp + \Avi\!,\,&&\Biamp + \Bvi), \label{eq:tap_av}\\
  \ci &= \fvc(\Aiamp + \Avi\!,\,&&\Biamp + \Bvi). \label{eq:tap_cv}
\end{alignat}

\section{Implementation}
Using the equations detailed in 
\eqref{eq:tap_AhBh}--\eqref{eq:tap_cv}, we can construct a fixed-point iteration
(FPI) which, given some arbitrary starting condition, can be
run until convergence in order to obtain the GRBM-inferred 
distribution on the signal variables defined by $\veca$, $\vecc$.
These distributions are then passed back to the CS observational
factors to complete the AMP iteration for CS reconstruction. We
detail this procedure in Alg. \ref{alg:grbmamp}. One important 
addition is the use of a damping step \cite{Hes2002} on the RBM-inferred 
values of $\veca$ and $\vecc$. We find that a fixed value of $\gamma = 0.5$, 
equally combining the previous and presently inferred moments, stabilizes
the interaction between the GRBM and the outer AMP inference. This becomes
especially important for $\alpha$ small, where oscillations between the 
two inference loops degrades reconstruction performance.

\begin{algorithm}[tb]
\small
    \caption{AMP with GRBM Signal Prior\label{alg:grbmamp}}  
  \begin{algorithmic}
    \STATE {\bfseries Input:} $\F$, $\mathbf{y}$, $\mathbf{W}$, 
                              $\prmtsVis$, $\prmtsHid$
    \STATE \emph{Initialize}: $\mathbf{a}$,$\mathbf{c}$,
                              ${\rm t} =1$
    
    \emph{Outer AMP Inference Loop:}
    \REPEAT
      \STATE AMP Update on $\{V_m,\omega_m\}$ as in \eqref{eq:ampV}, \eqref{eq:ampw}
      \STATE AMP Update on $\{\Aiamp,\Biamp\}$ as in \eqref{eq:ampA}, \eqref{eq:ampB}
        \STATE \emph{(Re)Initialize}:
            $\ai = \fva (\Aiamp, \Biamp) \; \forall i$,
            $\ahm = \chm = 0 \; \forall \mu$
        \emph{Inner RBM Inference Loop:}
        \REPEAT
          \STATE Update $\{\Avi,\Bvi\}$ 
            as in (\ref{eq:tap_AvBv})
          \STATE Update $\{\ai,\ci\}$
            as in (\ref{eq:tap_av}), (\ref{eq:tap_cv})
          \STATE Update $\{\Ahm,\Bhm\}$
            as in (\ref{eq:tap_AhBh})
          \STATE Update $\{\ahm,\chm\}$
            as in (\ref{eq:tap_ahch})
        \UNTIL{Convergence}        
        \STATE $\veca^{(\rm t)} = \gamma\cdot\veca^{(\rm t - 1)} 
                                     + (1-\gamma)\cdot\veca$
        \STATE $\vecc^{(\rm t)} = \gamma\cdot\vecc^{(\rm t - 1)} 
                                     + (1-\gamma)\cdot\vecc$
      \STATE ${\rm t} \gets {\rm t} + 1$
    \UNTIL{Convergence on $\mathbf{a}$}
  \end{algorithmic}
\end{algorithm}

While the Hamiltonian for the RBM model used here is in agreement
with the literature on real-valued RBMs, the TAP FPI on the RBM
points out one flaw in the construction of the real-valued
RBM model. 
Specifically, the unbounded nature of the energy for variables in $\real$
manifests in this context by
allowing \emph{negative} variance-like
terms $\Avi$ and $\Ahm$ which carry through into the prior-dependent
functions. 
We propose to handle this
dilemma of negative variance by forcing the truncation of $P_0(h_{\mu})$
and $P_0(x_i)$. 
Truncation can gracefully handle negative variances
by transforming these distributions to uniformity over the 
truncation bounds. In practice, truncation is an easy assumption to make on both the training data and the signal to be reconstructed. For example,
image data naturally lies within a specific range of values as defined by
the images' bit-depth. For sparse signals, such as those commonly studied in the
context of CS, we propose the use of a \emph{truncated} Gauss-Bernoulli 
distribution on the visible units, which only has non-zero probability density within
a fixed range.

\section{Experiments}

\begin{figure*}[ht!]  
  \centering
  \begin{minipage}{1\textwidth}
    \centering
    \includegraphics[width=0.6\textwidth]{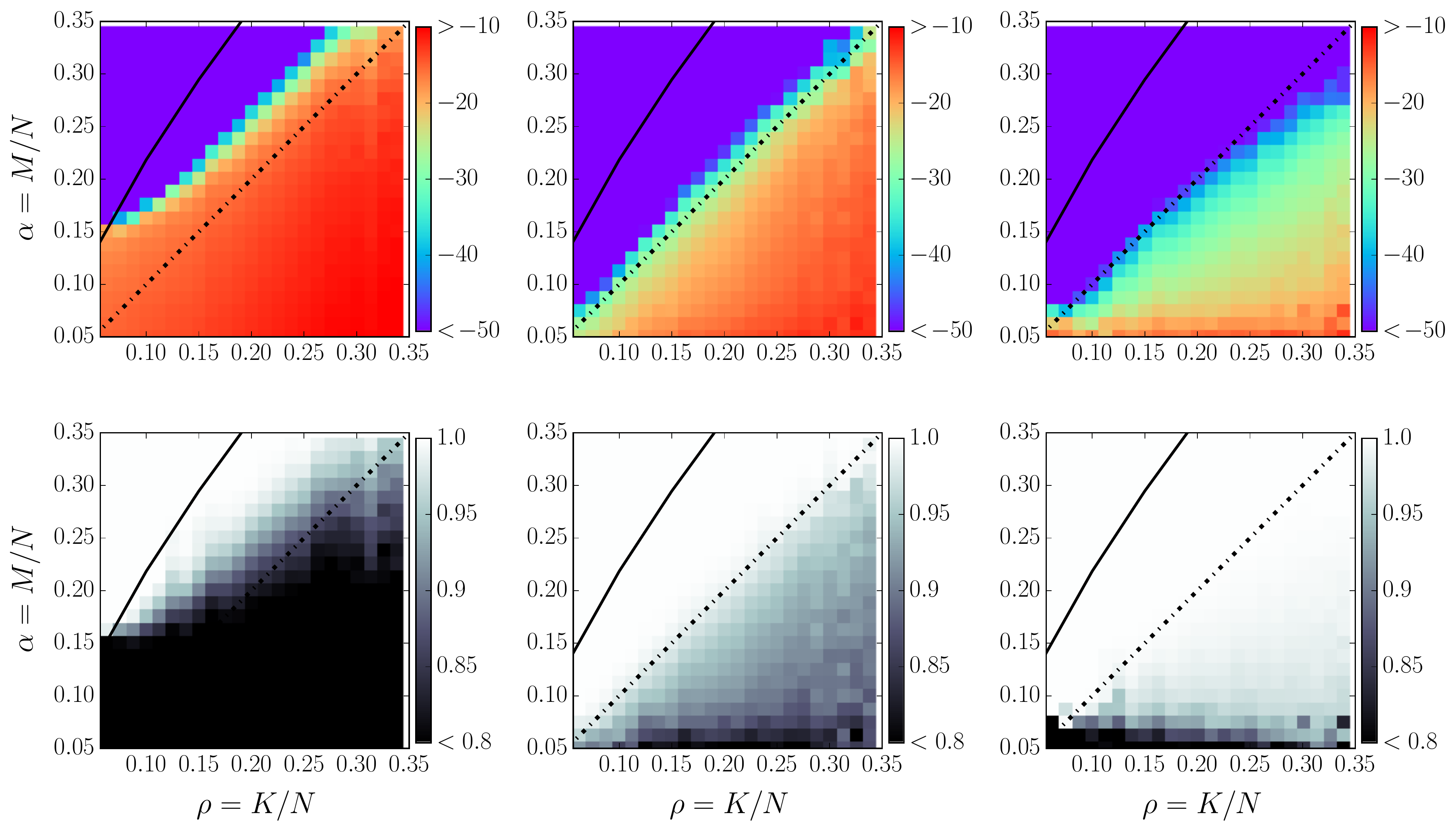}    
    ~~~~
    \raisebox{0.5cm}{\includegraphics[width=0.22\textwidth]{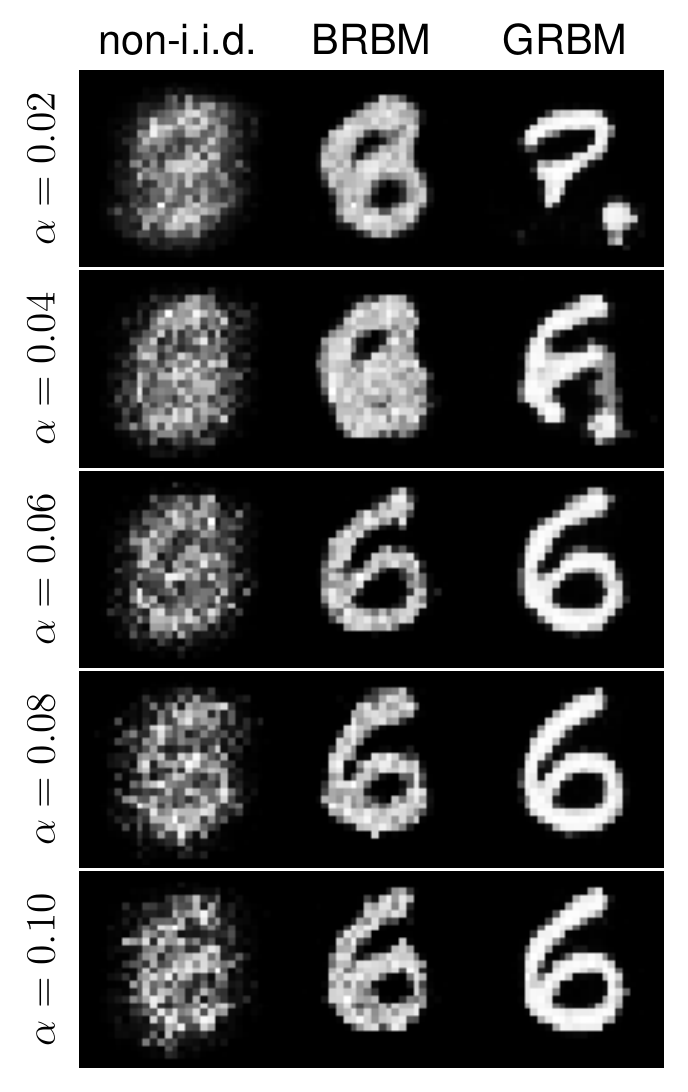}}
  \end{minipage}
  \caption{\label{fig:mnist}
    \textbf{(Left)}
    CS reconstruction performance over first 1,000 digit images from the
    MNIST test partition. Results for non-\iid AMP, support-based BRBM-AMP, and
    GRBM-AMP are on the left, center, and right, respectively. The $M=K$ oracle
    support transition is indicated by the black dotted line, and the spinodal
    transition \cite{KMS2012} by the solid one.
    \emph{Top:} Average reconstruction accuracy in MSE measured in dB.
    \emph{Bottom:} Average reconstruction correlation with original digit
     image.
    \textbf{(Right)} 
    Visual comparison of reconstructions for a single digit image 
    ($\rho = 0.25$) for small values of $\alpha$. 
  }
\end{figure*}

We now present the results of our numerical studies of 
GRBM-AMP performance for the AWGN CS reconstruction task. 
For all reconstruction 
tasks, an AWGN noise variance $\Delta = 10^{-8}$ was 
used. Additionally, all elements of the sampling matrices $\F$ 
were  drawn from a zero-mean Gaussian distribution of variance
$1/\sqrt{N}$. 
 
The results we present are based on the MNIST handwritten
digit dataset \cite{LBB1998} which consists of $28\times28$ gray-scale digit 
images split between 60,000 training samples and 10,000 test samples. While
we train over the entire MNIST training partition, we conduct our
reconstruction experiments for the first 1,000 digit images drawn from the test 
partition. We test three different approaches for this dataset. The first, 
termed non-\iid AMP, consists of empirically estimating the per-coefficient 
prior hyper-parameters from the training data. This approach assumes a fully
factorized model of the data, neglecting any covariance structure between the 
coefficients. The second approach is that of \cite{tdk2016}, here termed 
binary-support RBM (BRBM-AMP), which uses a binary RBM to model the correlation
structure of the support, alone. Finally, we test the proposed GRBM-AMP, 
using a general RBM trained with binary hidden units and 
Gauss-Bernoulli visible units, which models the data in its ambient domain.

To train the GRBM parameters for MNIST, we use a GRBM with 784 binary 
hidden units.
The RBM can be trained using either contrastive divergence, sampling 
the visible units from a truncated GB prior, or using the 
GRBM TAP iteration shown here in conjunction with the EMF strategy of 
\cite{GTZ2015}, a strategy which we detail in a forthcoming work. For the 
specific GRBM model we use for these CS experiments, we train a GRBM using the
EMF approach for 150 epochs using a learning rate of 0.01 with an $\ell_2$
weight decay penalty of 0.001. Learning momentum of 0.5 was used.
Finally, the truncation bounds were set to the range $[0,1]$.
In the case of the BRBM, where all variables are binary, EMF training
\cite{GTZ2015} was used with a learning rate of 0.005, while other
parameters have been set the same as for the GRBM.

We present the results of the three approaches in Fig. \ref{fig:mnist}, where
we evaluate reconstruction performance over the test set in terms of both 
MSE, measured in decibels, and correlation between the reconstructed and 
original images, where correlation is measured as 
$(\x-\bar{x})^T(\veca-\bar{a})/\sigma_{\x}\sigma_{\veca}$. 
In both of these comparisons, we show performance over the 
phase diagram, where $\alpha$ refers to the number of CS measurements observed
and $\rho$ refers to the overall sparsity for each digit image, $K/N$, where
$K$ is the number of non-zero pixels in the digit image. As many of the tested
images posses few non-zeros, the test dataset is skewed towards small $\rho$,
hence the increased variability at $\rho > 0.3$ for Fig. \ref{fig:mnist}. From 
these results, we can see a clear progression of reconstruction performance as
we move from an empirical factorized model, to a model of the support alone,
to a model of the signal itself. For non-\iid AMP, we see that the localized
information from the training set allows for a transition curve which is 
parallel to $M=K$, which is in contrast to the well-known transition for 
AMP with an \iid GB signal prior. For BRBM-AMP we observe that the reconstruction
transition lies very close to the $M=K$ optimal line, as also observed in
\cite{tdk2016}, showing the advantage of
leveraging the strong support correlations which exist in the dataset.
In the case of the GRBM-AMP, we observe that
the maximal performance is no longer bounded by the $M=K$ line, with 
low MSE achievable even for $M<K$. This is an
intuitive result, as the GRBM model provides information not only about the 
support, but also about the values of the signal on that support. 
This effect is most drastically observed in terms of the average reconstruction
correlation, where we observe 
almost perfect correlation for the entire test set for $\alpha > 0.10$, 
independent of the signal sparsity.

Finally, we also see a visual comparison for one example of a reconstructed digit
in Fig. \ref{fig:mnist} in the small $\alpha$ setting. In this extreme case, we 
can see that both the BRBM- and GRBM-AMP produce reconstructions whose support
closely match the original `6' digit image. We note that the GRBM is able to 
capture the smoothness of the digit image where the BRBM cannot.

\section{Conclusion}
In this work, we derived an AMP-based algorithm for CS signal reconstruction 
using an RBM to model the signal class in its ambient domain. To accomplish
this modeling, we developed a model for a class of general RBMs, allowing for
arbitrary distributions on the hidden and visible units. To allow the use of 
such a model within AMP, we proposed a TAP-based approximation of the RBM which
we derived from belief propagation. By performing inference on the RBM under the
influence of the outer AMP inference, we have developed
a novel algorithm for CS reconstruction of sparse structured data. The proposed 
approach can be of great use in signal reconstruction contexts where there 
exists an abundance of data which lack developed correlation models. Additionally,
the inference we propose for general RBMs could be used to develop novel 
generative models by varying its architecture and the distribution of 
the hidden variables.

{\small
\section*{Acknowledgment}
This research was funded by European Research Council under the European Union’s 7th Framework Programme (FP/2007-2013/ERC Grant Agreement 307087-SPARCS).}

\bibliographystyle{IEEEtran}
\bibliography{references}

\begin{thebibliography}{10}
\providecommand{\url}[1]{#1}
\csname url@samestyle\endcsname
\providecommand{\newblock}{\relax}
\providecommand{\bibinfo}[2]{#2}
\providecommand{\BIBentrySTDinterwordspacing}{\spaceskip=0pt\relax}
\providecommand{\BIBentryALTinterwordstretchfactor}{4}
\providecommand{\BIBentryALTinterwordspacing}{\spaceskip=\fontdimen2\font plus
\BIBentryALTinterwordstretchfactor\fontdimen3\font minus
  \fontdimen4\font\relax}
\providecommand{\BIBforeignlanguage}[2]{{%
\expandafter\ifx\csname l@#1\endcsname\relax
\typeout{** WARNING: IEEEtran.bst: No hyphenation pattern has been}%
\typeout{** loaded for the language `#1'. Using the pattern for}%
\typeout{** the default language instead.}%
\else
\language=\csname l@#1\endcsname
\fi
#2}}
\providecommand{\BIBdecl}{\relax}
\BIBdecl

\bibitem{CW2008}
E.~J. Cand{\`e}s and M.~B. Wakin, ``An introduction to compressive sampling,''
  \emph{{IEEE} Signal Processing Magazine}, vol.~25, no.~2, pp. 21--30, 2008.

\bibitem{CR2005a}
E.~Cand\`{e}s and J.~Romberg, ``Signal recovery from random projections,'' in
  \emph{Computational Imaging III}.\hskip 1em plus 0.5em minus 0.4em\relax
  Proc.~SPIE 5674, Mar. 2005, pp. 76--86.

\bibitem{Don2006}
D.~L. Donoho, ``Compressed sensing,'' \emph{{IEEE} Trans. on Information
  Theory}, vol.~52, no.~4, pp. 1289--1306, Apr. 2006.

\bibitem{TLW2006}
D.~Takhar, J.~N. Laska, M.~B. Wakin, M.~F. Duarte, D.~Baron, S.~Sarvotham,
  K.~F. Kelly, and R.~G. Baraniuk, ``A new compressive imaging camera
  architecture using optical-domain compression,'' in \emph{{Computational
  Imaging IV}}.\hskip 1em plus 0.5em minus 0.4em\relax Proc.~SPIE 6065, 2006,
  p. 606509.

\bibitem{MED2011}
M.~Mishali, Y.~C. Eldar, O.~Dounaevsky, and E.~Shoshan, ``Xampling: Analog to
  digital at sub-{Nyquist} rates,'' \emph{IET Circuits, Devices and Systems},
  vol.~5, no.~1, pp. 8--20, January 2011.

\bibitem{ZXY2015}
Z.~Zhang, Y.~Xu, J.~Yang, X.~Li, and D.~Zhang, ``A survey of sparse
  representation: Algorithms and applications,'' \emph{{IEEE} Access}, vol.~3,
  pp. 290--530, 2015.

\bibitem{DTD2006}
D.~L. Donoho, Y.~Tsaig, I.~Drori, and J.-L. Starck, ``Sparse solution of
  underdetermined linear equations by stagewise orthogonal matching pursuit,''
  Stanford University, Tech. Rep., 2006.

\bibitem{DGN2008}
T.~T. Do, L.~Gan, N.~Nguyen, and T.~D. Tran, ``Sparsity adaptive matching
  pursuit algorithm for practical compressed sensing,'' in \emph{Proc.
  {Asilomar} Conf. on Signals, Systems, and Computers}, Pacific Grove,
  California, Oct. 2008, pp. 581--587.

\bibitem{DT2006}
D.~L. Donoho and J.~Tanner, ``Thresholds for the recovery of sparse solutions
  via l1 minimization,'' in \emph{Information Sciences and Systems, Proc.
  Annual Conference on}, 2006, pp. 202--206.

\bibitem{BSB2010}
D.~Baron, S.~Sarvotham, and R.~G. Baraniuk, ``Bayesian compressive sensing via
  belief propagation,'' \emph{{IEEE} Trans. on Signal Processing}, vol.~58,
  no.~1, pp. 269--280, 2009.

\bibitem{Ran2010}
S.~Rangan, ``Estimation with random linear mixing, belief propagation and
  compressed sensing,'' in \emph{Proc. Annual Conf. on Information Sciences and
  Systems}, 2010, pp. 1--6.

\bibitem{DMM2009}
D.~L. Donoho, A.~Maleki, and A.~Montanari, ``Message-passing algorithms for
  compressed sensing,'' \emph{Proc. Nat. Academy of Sciences of the U.S.A.},
  vol. 106, no.~45, p. 18914, 2009.

\bibitem{DMM2010b}
------, ``Message passing algorithms for compressed sensing: {II}. analysis and
  validation,'' in \emph{Proc. {IEEE} Info. Theory Workshop}, Cairo, Egypt,
  2010, pp. 1--5.

\bibitem{Ran2011}
S.~Rangan, ``Generalized approximate message passing for estimation with random
  linear mixing,'' in \emph{Proc. {IEEE} Intl. Symp. on Info. Theory}, 2011, p.
  2168.

\bibitem{KMS2012}
F.~Krzakala, M.~M{\'e}zard, F.~Sausset, Y.~Sun, and L.~Zdeborov{\'a},
  ``Probabilistic reconstruction in compressed sensing: Algorithms, phase
  diagrams, and threshold achieving matrices,'' \emph{Journal of Statistical
  Mechanics: Theory and Experiment}, vol. 2012, no.~8, p. P08009, 2012.

\bibitem{KrzakalaPRX2012}
------, ``Statistical physics-based reconstruction in compressed sensing,''
  \emph{Phys. Rev. X}, vol.~2, p. 021005, 2012.

\bibitem{dremeau2012boltzmann}
A.~Dr{\'e}meau, C.~Herzet, and L.~Daudet, ``{B}oltzmann machine and mean-field
  approximation for structured sparse decompositions,'' \emph{{IEEE} Trans. on
  Signal Processing}, vol.~60, no.~7, pp. 3425--3438, 2012.

\bibitem{tdk2016}
E.~W. Tramel, A.~Dr{\'e}meau, and F.~Krzakala, ``Approximate message passing
  with restricted {B}oltzmann machine priors,'' \emph{Journal of Statistical
  Mechanics: Theory and Experiment}, 2016, to appear.

\bibitem{KLT2016}
K.~Kulkarni, S.~Lohit, P.~Turaga, R.~Kerviche, and A.~Ashok, ``{ReconNet}:
  Non-iterative reconstruction of images from compressively sensed random
  measurements,'' 2016, arXiv:1601.06892.

\bibitem{MPB2015}
A.~Mousavi, A.~B. Patel, and R.~G. Baraniuk, ``A deep learning approach to
  structured signal recovery,'' 2015, arXiv:1508.04065.

\bibitem{KM2015}
U.~S. Kamilov and H.~Mansour, ``Learning optimal nonlinearities for iterative
  thresholding algorithms,'' 2015, arXiv:1512.04754.

\bibitem{schniter2010turbo}
P.~Schniter, ``Turbo reconstruction of structured sparse signals,'' in
  \emph{Proc. Conf. on Info. Sciences and Systems}, 2010, pp. 1--6.

\bibitem{rangan2011hybrid}
S.~Rangan, A.~K. Fletcher, V.~K. Goyal, and P.~Schniter, ``Hybrid approximate
  message passing with applications to structured sparsity,'' 2011,
  arXiv:1111.2581.

\bibitem{GTZ2015}
M.~Gabri{\'e}, E.~W. Tramel, and F.~Krzakala, ``Training restricted {B}oltzmann
  machines via the {Thouless-Andreson-Palmer} free energy,'' in \emph{Advances
  in Neural Information Processing System}, vol.~28, Montreal, Canada, June
  2015, pp. 640--648.

\bibitem{Smo1986}
P.~Smolensky, \emph{Information Processing in Dynamical Systems: Foundations of
  Harmony Theory}, ser. Parallel Distributed Porcessing: Explorations in the
  Microsctructure of Cognition.\hskip 1em plus 0.5em minus 0.4em\relax MIT
  Press, 1986, ch.~6, pp. 194--281.

\bibitem{Hin2002}
G.~E. Hinton, ``Training products of experts by minimizing contrastive
  divergence,'' \emph{Neural Computation}, vol.~14, no.~8, pp. 1771--1800,
  2002.

\bibitem{HT2015}
H.~Huang and T.~Toyoizumi, ``Advanced mean-field theory of restricted
  {Boltzmann} machine,'' \emph{Physical Review E}, vol.~91, no.~5, p. 050101,
  2015.

\bibitem{Hes2002}
T.~Heskes, ``Stable fixed points of loopy belief propagation are minima of the
  {Bethe} free energy,'' in \emph{Advances in Neural Information Processing
  Systems}, vol.~15, 2002, pp. 359--366.

\bibitem{LBB1998}
Y.~LeCun, L.~Bottu, Y.~Bengio, and P.~Haffner, ``Gradient-based learning
  applied to document recognition,'' \emph{Proc. of the {IEEE}}, vol.~86,
  no.~11, pp. 2278--2324, November 1998.

\end{thebibliography}

\end{document}